\documentclass[a4paper,10pt]{article}
\newcommand{\head}[1]{\textnormal{\textbf{#1}}}
\usepackage[onehalfspacing]{setspace}
\usepackage{amsmath}
\usepackage{array}
\usepackage{booktabs}
\usepackage{multirow}
\newcommand{\normal}[1]{\multicolumn{1}{l}{#1}}
\usepackage{amssymb}
\usepackage{hyperref}
\usepackage{fontenc}
\usepackage{inputenc}
\usepackage{authblk}
\usepackage{color}
\usepackage{graphicx}
\usepackage{cite}
\usepackage{slashed}
\hypersetup{colorlinks=true,linkcolor=magenta,linktoc=page,citecolor=blue}
\usepackage{tikz}
\usepackage{multirow}
\usepackage{hhline}
\usepackage[top=1.5cm,bottom=2.5cm,right=2.54cm,left=2.54cm]{geometry}
\definecolor{c1}{rgb}{0.5,0,1}
\colorlet{aqua}{-red!75}
\newcommand{\be}{\begin{equation}}
\newcommand{\bea}{\begin{eqnarray}}
\newcommand{\eea}{\end{eqnarray}}
\newcommand{\ba}{\begin{array}}
\newcommand{\ea}{\end{array}}
\newcommand{\ee}{\end{equation}}
\numberwithin{equation}{section}

\begin{document}
\begin{flushright}
 IPM/P-2017/063 \\
\end{flushright}

\vspace*{20mm}
\begin{center}
{\Large {\bf On universal terms of holographic entanglement entropy in theories with hyperscaling violation }\\}

\vspace*{15mm} \vspace*{1mm} {M. Reza Tanhayi}

 \vspace*{1cm}

{Department of Physics, Faculty of Basic Science, Islamic Azad
University Central Tehran Branch (IAUCTB), P.O. Box 14676-86831,
Tehran, Iran\\
School of Physics, Institute for Research in Fundamental Sciences
(IPM) P.O. Box 19395-5531, Tehran, Iran}

 \vspace*{0.5cm}
{E-mail: {\tt mtanhayi@ipm.ir}}

\vspace*{1cm}
\end{center}

\begin{abstract}

We examine the holographic entanglement entropy in hyperscaling violating backgrounds. Precisely in such theories by semi-analytic computation, we use  holographic methods to derive the universal terms of entanglement entropy in various dimensions. We also find such terms when the bulk action includes higher-order curvature terms. We make further comments on the mutual information in  hyperscaling violating backgrounds with higher curvature interactions. We explore the sign of tripartite information which results in a monogamy property on mutual information. It is shown that for various hyperscaling parameters, the mutual and tripartite information are always positive and negative, respectively. This indicates that mutual information is monogamous in theories with hyperscaling violation.   

\end{abstract}
\newpage

\section{Introduction}

The gauge/gravity correspondence \cite{Maldacena:1997re}
claims an equivalence between strongly coupled  gauge theory and gravity theory which lives in one higher dimensional Anti-de Sitter (AdS) spacetime. In this correspondence the objects in the gravitational theory are related to  those in the dual
field theory and in this sense it is known as the duality. This duality could actually provide a framework to study quantum physics of strongly correlated many-body systems via the classical dynamics of gravity \cite{Hartnoll:2009sz}. As a matter of fact, AdS geometries are dual to the field theory with conformal symmetry; However, studying the field theories which are scale-invariant but not conformal invariant and finding the corresponding dual metrics which are not asymptotically AdS can be a challenging task. The main motivation for studying such generalization comes from this fact that many physical systems in their critical points exhibit a rather different scaling in space and time, thus finding the holographic dual models for such systems could be a relevant question. For example in addressing the Landau-Fermi liquids \cite{Hartnoll:2009sz, Ogawa:2011bz}, one needs a Lifshitz type metrics in dual gravity theory  where the spatial and time coordinates of the field theory have been scaled differently \cite{Taylor:2008tg}. The class of dual geometries with such scaling may be written as 
\be\label{lif1}
ds^2_{d+2} =\tilde{L}^2\Big(\frac{-dt^2}{r^{2z}}
+\frac{dr^2}{r^2} +\frac{1}{r^2}\sum_{i=1}^ddx_i^2\Big),\ee
where  $z$ is the dynamical critical exponent and $\tilde{L}$ stands for the curvature scale. The above metric reduces to AdS$_{d+2}$ for $z=1$ and enjoys the following scaling of the coordinates 
 \be\label{scale}
t\rightarrow \zeta^z
t,\,\,\,\,\,\,\,\,\,\,\,\,\,\vec{x}\rightarrow\zeta\vec{x},\,\,\,\,\,\,\,\,\,\,\,\,\,r\rightarrow
\zeta r.\ee  Note that the metric \eqref{lif1} arises as the solution of simple gravity theories coupled with some sort of matter field \cite{Taylor:2008tg, Kachru:2008yh}.  One may also add a scalar dilaton field in the bulk theory to obtain a larger class of scaling metrics which has been studied in several works \cite{Dong:2012se}. On the other hand studying both dilaton scalar field together with an abelian gauge field leads to a full class of scaling metrics named as the hyperscaling violating geometries \cite{Charmousis:2010zz} \be\label{lif2}
 ds^2_{d+2} =\tilde{L}^2r^{\frac{2\theta}{d}} \Big(\frac{-dt^2}{r^{2z}}
 +\frac{dr^2}{r^2} +\frac{1}{r^2}\sum_{i=1}^ddx_i^2\Big),\ee
where $\theta $ is the hyperscaling violating exponent. Note that one recovers  a non-relativistic scaling symmetry namely the Lifshitz geometry by setting $\theta=0$ and $z\neq 1$. It is clear that the metric \eqref{lif2} can not be an ordinary solution of the Einstein equation and in fact the isotropy should be broken by a gauge field or a massive vector field coupled to a scalar field
\cite{Kachru:2008yh, Dong:2012se, Balasubramanian:2008dm, Charmousis:2010zz, Alishahiha:2012qu,
Roychowdhury:2015fxf}. This metric under the
scale-transformation \eqref{scale} transforms as
$ds\rightarrow\zeta^{\frac{\theta}{d}}ds$.\\ It was shown that in the  $d+1$ dimensional field theories dual to the hyperscaling violating metric \eqref{lif2} the corresponding thermodynamical parameters say as entropy scale with respect to temperature as  $T^{(d-\theta)/z}$
\cite{Charmousis:2010zz, Gouteraux:2011ce}, on the other hand in the Lifshitz metric the entropy scales as $T^{d/z}$ which indicates by turning on the hyperscaling violating parameter the degrees of freedom effectively can be described in  $d-\theta$ dimensions \cite{such}. Remarkably, by associating an effective dimension to such theories, one obtains the log behavior of the entanglement entropy. This could be utilized in a holographic description of the systems with Fermi surface in condensed matter physics  \cite{Alishahiha:2012cm}, namely in any
dimensions one can engineer the parameters to provide a gravitational dual for a system with Fermi
surface

The appeared parameters in metric \eqref{lif2} are constrained by imposing some conditions e.g., in Ref. \cite{Dong:2012se} it was shown that holographic consideration of the null energy condition yields \begin{eqnarray}
(d-\theta)(z-1-\frac{\theta}{d})\ge0,\hspace{1cm}(z-1)(d+z-\theta)\ge0,
\end{eqnarray}
where we will consider $0\le\theta<d$. 

In this work we will further investigate some specific features of hyperscaling violating theories by making use of the gravity dual which is described by Einstein or Gauss-Bonnet gravity coupled to some specific matter fields. Precisely, we deal with the structure of entanglement in the vacuum of such theories by computing the holographic entanglement entropy for a strip, a sphere and a cylinder as the entangling regions. In the case of a strip, an explicit form of the entanglement entropy is computed, whereas the logarithmically divergent terms are studied for a spherical or a cylindrical entangling surfaces.\\ It is known that entanglement entropy for local quantum field theories is infinite and generally, for a given entangling region with area $\mathcal{A}$ and in a $d+1$ dimensional field theory one obtains \cite{Srednicki:1993im-Casini:2006hu}

\be S=c_{d-1} \frac{\mathcal{A}}{\epsilon^{d-1}}+\cdots+\mbox{s}_{d+1}+{\cal O}(\frac{\epsilon}{\ell}),\,\,\,\,\,\,\,\,\mbox{s}_{d+1}=\left\{ {\begin{array}{*{20}{c}}
		c_{d+1}\log\frac{\ell}{\epsilon}, \hfill & {\mbox{for even-dimensional CFTs}}, \hfill \\
		c'_{d+1}, \hfill & {\mbox{for odd-dimensional CFTs}}\hfill  \\
	\end{array}} \right.\ee
where $c_i$'s are some constants, $\epsilon$ is the UV cutoff and $\ell$ stands for the characteristic length of corresponding entangling region. The most divergence term is proportional to the area of the entangling surface and this is known as the area law which is due to the infinite correlations between degrees of freedom near the boundary of entangling surface. $s_{d+1}$ is the universal part of the entanglement entropy and for even-dimensional CFTs (odd $d$ in our notation), there appears a logarithmic term where its coefficient is physical and universal in a sense that it is not affected by cutoff redefinitions. It is argued that it could carry some universal information about the underlying quantum field theory and thus is of great interest.  For example, the entanglement entropy of a strip of width $\ell$  in two dimensions becomes \cite{Nishioka:2009un}
\be S=\frac{c}{3}\log\frac{\ell}{\epsilon},\ee
where its coefficient is indeed the central charge of the theory. On the other hand for spherical and cylindrical entangling regions there are also logarithmic terms which the corresponding coefficients could potentially address the anomalies of underlying quantum field theory. It is known that in four dimensional conformal theory the coefficients of these logarithmic terms are related to $a$ and $c$ central charges, \cite{Solodukhin:2008dh}  and notably for higher dimensional field theories not much is known about the universal logarithmic term of entanglement entropy \cite{Hung:2011xb-Myers:2012ed-Safdi:2012sn}.

In this paper, we investigate the logarithmically divergent terms of Holographic Entanglement Entropy (HEE) and by semi-analytic consideration we find the logarithmic terms for some values of the parameters in hyperscaling violating theories. We use holographic description of entanglement entropy for both Einstein gravity and higher order gravity theories for three entangling regions, strip, sphere and cylinder in theories with hyperscaling violation. 

It is worth noting, however, that the entanglement entropy is a scheme-dependent quantity due to the appearance of UV cutoff parameter. Moreover the entanglement entropy cannot demonstrate all the physical content of a field theory, for example as mentioned above, for a 2-dimensional conformal field theory entanglement entropy for a single interval is completely fixed by the central charge of the theory. Thus to investigate the other aspects of conformal field theory and  improve our knowledge of the Hilbert space of a quantum system, it seems some other entanglement measures are needed. One of the most powerful of such measures is indeed the mutual information which is a UV cutoff-independent quantity \cite{Casini:2008wt}. We also study the mutual information and its sign in hyperscaling violating theories.  


The organization of the paper is as follows. In the next section we will consider the holographic entanglement entropy for strip, spherical and cylindrical entangling regions in hyperscaling violating theories. In section 3 we will explore the curvature corrections to holographic entanglement entropy and also the universal terms in higher curvature hyperscaling gravity theories will be addressed in this section. We also make further
comments on mutual information and its monogamy property for a trip geometry  in section 4.  Last section is devoted to discussions.

\section{Holographic entanglement entropy in Hyperscaling violating backgrounds: Einstein gravity }

Let us suppose a system with Hilbert space ${\cal H}$ which has been divided in two parts $A$ (mostly
supposed as a spatial region) and its complement $\bar{A}$ then ${\cal H}={\cal H}_A \otimes {\cal H}_{\bar{A}}$. In principle, entanglement entropy can be used as a measure of quantum correlation between subsystems $A$ and $\bar{A}$ where the corresponding entanglement entropy for subsystem $A$ is given by the  von Neumann entropy as follows \cite{Calabrese:2004eu, Horodecki:2009zz} 
\be
 S_A=-{\rm Tr}(\rho_A
 \log\rho_A).\ee
The reduced density matrix $\rho_A$ can be obtained by tracing over the degrees of freedom living in the environment of $A$ namely in  $\bar{A}$.  The entanglement entropy of $A$ depends on the chosen region and also on the state of the field theory. Basically, computing the entanglement entropy is a difficult task, but, in a field theory with a holographic dual described by classical Einstein gravity, there is an elegant proposal to compute this quantity made by Ryu and Takayanagi known as RT- proposal \cite{Nishioka:2009un}. According to this proposal, in a given entangling region, the entanglement entropy is given by 
\be S=\frac{{\cal
		A}}{4G_N},\ee
 where $G_N$ is the Newton constant and ${\cal
	A}$ is  minimal surface in the bulk whose
boundary coincides with the boundary of the entangling region. This is indeed the holographic description of the entanglement entropy. 

In this section we study the holographic entanglement entropy for systems with hyperscaling violation. It was
shown that an Einstein-Maxwell-dilaton model with a particular potential which is given by the following action
\be
{\cal I}=\frac{1}{16\pi G_N}\int d^{d+2}x\,\sqrt{-g}\Big[R-\frac{1}{2}(\partial\phi)^2+V_0e^{\gamma\phi}-\frac{1}{4}e^{\alpha\,\phi}F^{\mu\nu}F_{\mu\nu}\Big],
\ee
admits a solution of the form \eqref{lif2}, where $\gamma$, $V_0$, and $\alpha$ are some free parameters of the model. The gauge field produces an anisotropic scaling
symmetry of the theory and a hyperscaling violating factor comes from  the above particular form of the potential. The scalar and gauge fields are given by
\be \phi=\phi_0+\beta\ln r,\,\,\,\,\,\,\,\,
A_t=\sqrt{\frac{2(z-1)}{d-\theta+z}}e^{\frac{d-1+\theta/d}{\beta}\phi_0}r^{d-\theta+z},
\ee
where $\beta=\sqrt{2(d-\theta)(z-1-\theta/d)}$. Noting that the free parameters of the action are fixed by the hyperscaling violating and dynamical exponent parameters as follows \cite{Alishahiha:2012qu}
\be V_0=\frac{(d-\theta+z-1)(d-\theta+z)}{e^{\gamma\phi_0}},\,\,\,\,\lambda=\frac{2\theta(d-1)-2d^2}{\beta d},\,\,\,\gamma=\frac{2\theta}{\beta d}.\ee 
In the hyperscaling violating background, we first review the holographic entanglement entropy for a strip then generalize the consideration to  spherical and  cylindrical entangling surfaces by computing the logarithmically divergent terms.

\subsection{Strip entangling region}
 First let us review the holographic entanglement entropy for a  strip at a fixed time on the boundary which is defined by  \be \label{region}t=\mbox{fixed},\,\,\,-\frac{\ell}{2}\leq
 x_1\leq \frac{\ell}{2},\hspace{1cm}0\leq x_2,\cdots,x_d\leq H,\ee where $(t, \vec{x})$ are the space-time
 coordinates. Now the aim is to find the surface in the bulk with
 the boundary of the above strip and then minimize it. In the metric \eqref{lif2} the profile of
 the hypersurface in the bulk may be parameterized by $x(r)$, thus the induced metric on this hypersurface is
 given by\footnote{In this section we set $\tilde{L}$ to one.} \be
 ds_{\mbox{ind}}^2=r^{2\frac{\theta}{d}-2}\bigg[\Big(1+x'^2\Big)dr^2+\sum_{a=2}^d dx_a^2\bigg]\ee
 where here the prime stands for the derivative with respect to
 $r$. Thus the area of the profile in the bulk is given by 
 \be \label{area1}{\cal A}=\int
\sqrt{\sigma} d^dx=\frac{H^{d-1}}{2}\int
dr{(r^{\theta-d})}{\sqrt{1+{x'}_1^2}},\ee
where $\sigma$ stands for the induced metric determinant. After minimizing the functional of \eqref{area1} and using the condition of the hypersurface turning point $r_t$, in which  $x_1'(r_t)\rightarrow \infty$, one gets the following conserved quantity along the radial profile
\begin{equation}\label{eom}
-\frac{r^{\theta -d} x_1'}{\sqrt{x_1'^2+1}}=-r_t^{\theta-d},.
\end{equation}
One can use the above relation to find the width of the strip as a function of turning point and for $\theta\neq d-1$ one obtains
\begin{equation}\label{ell}
\ell=2\int_{0}^{r_t}x_1'(r)\,dr=\frac{2\sqrt{\pi}\Gamma(\frac{1+d-\theta}{2(d-\theta)})}{\Gamma(\frac{1}{2(d-\theta)})}r_t.
\end{equation}
This can be inverted to find the turning point of the proposed hypersurface in the bulk and finally the HEE for Einstein Gravity (EG) reads as follows 
\begin{equation}\label{EE}
S_{EG} = \frac{{{H^{d-1}}}}{{2{G_N(d-\theta-1)}}}\left(\frac{1}{{{\epsilon ^{(d-\theta-1)}}}}- \left(\frac{\sqrt{\pi}\Gamma(\frac{1+d-\theta}{2(d-\theta)})}{\Gamma(\frac{1}{2(d-\theta)})}\right)^{d-\theta
	}(\frac{\ell}{2})^{\theta-d+1}\right),
\end{equation}
where $\epsilon$ stands for the UV scale which has been defined by the radial profile. This was found in \cite{Dong:2012se} and it is clear that the scaling behavior of HEE has been modified by an additional term as $(length)^\theta$. Noting that $\theta=0$ leads to the HEE for AdS$_{d+2}$ background.
\subsection{Sphere entangling region}
The holographic entanglement entropy for a spherical entangling region has been studied in the literature \cite{Nishioka:2009un}. Here, after briefly summarizing the results, we generalize the study to the hyperscaling violating backgrounds.  In this case, let us modify the metric \eqref{lif2} as $\sum dx^2_i=d\rho^2+\rho^2d\Omega_{d-1}^2$. On the boundary, the entangling region is a sphere with radius $r<R$, therefore, the corresponding co-dimension two hypersurface in the bulk is realized by $t=0$ and $\rho=\rho(r)$. With this assumption, the induced metric becomes

\be\label{lif2ind}
ds^2_{ind} =r^{\frac{2\theta}{d}-2} \Big((1+\rho'^2)dr^2+\rho^2d\Omega_{d-1}^2
\Big),\ee
the corresponding entropy functional reads as
\be\label{stripee}
S=\frac{{\cal A}_{d-1}}{4G_N}\int dr \rho^{d-1} r^{ \theta -d} \sqrt{ \rho'^2+1},
\ee
where ${\cal A}_{d-1}=\frac{2\pi^{d/2}}{\Gamma(\frac{d}{2})}$ is indeed the area of the unit $d-1$ dimensional sphere. The above  entropy functional should be minimized which leads to the following equation of motion
\be
\left(\rho '(r)^2+1\right) \Big((\theta -d) \rho (r) \rho '(r)-(d-1) r\Big)+r \rho (r)
\rho ''(r)=0.\label{profile}
\ee
 For  $\theta=0$ it has a simple solution as follows
\be \rho(r)=\sqrt{R^2-r^2},\ee
 and the corresponding  HEE is given by \cite{Nishioka:2009un}
  \be
S=\frac{\pi^{d/2}}{2G_N\Gamma(\frac{d}{2})}\bigg[{\cal P}_1(\frac{R}{\epsilon})^{d-1}+{\cal P}_3(\frac{R}{\epsilon})^{d-3}+\cdots+\left\{ {\begin{array}{*{20}{c}}
		{{\cal P}_{d-1}(\frac{R}{\epsilon})+{\cal
				 P}_d}, \hfill & {\mbox{for even $d$}}, \hfill \\
		{\cal P}_{d-2}(\frac{R}{\epsilon})^2+{\cal Q}\log(\frac{R}{\epsilon}), \hfill & {\mbox{for odd $d$}}. \hfill  \\
	\end{array}} \right.\bigg]\ee

\be\label{takaya}\nonumber {\cal P}_1=\frac{1}{d-1},{\cal P}_3=-\frac{d-2}{2(d-3)},\,\,{\cal P}_d=\frac{\Gamma(\frac{d}{2})}{2\sqrt{\pi}}\Gamma(\frac{1-d}{2}),\,{\cal Q}=(-1)^{\frac{d-1}{2}}\frac{(d-2)!!}{(d-1)!!}\ee
The coefficient of leading UV divergent term is proportional to the area of the boundary which is the area
law. On the other hand, when $d$ is odd, there is a logarithmic term and its coefficient is universal in a sense that it is independent of the rescaling.

The next step is to investigate the effect of dynamical and hyperscaling violation parameters on the universal term of HEE. However, it is important to note that to
identify the universal contribution, the near boundary behavior of the minimal surface would be sufficient. Thus, one needs to consider the asymptotic behavior of the profile near the boundary where can be written as
\be \rho(r)=R+c_1 r +c_2 r^2+c_3 r^3+\cdots.\ee
Taking into account the condition $\rho(0)=R$ and after minimizing the entropy functional one finds
\be c_1=0,\,\,\,\,c_2=\frac{(d-1) }{2R (\theta-d +1)},\,\,\,c_3=0.\ee
Thus near the boundary,  one obtains
\be
 S=\frac{1}{2G_N}\frac{\pi^{d/2}}{\Gamma(\frac{d}{2})}\int dr\Big[r^{\theta-d}\Big(R^{d-1}+\frac{(d-1)^2R^{d-3}(\theta-d+2)}{2(\theta-d+1)^2}r^2\Big)+{\cal O}(r^3)\Big].
 \ee
By semi-analytic computation we collect universal $\log$ terms in table 1.
\begin{table}[h!]

	\vspace{10pt}
	\centering
\begin{tabular}{c|cccccc}
	\hline
	   & $d=2$ & $d=3$ & $d=4$ & $d=5$ & $d=6$&$d=7$  \\
	\hline
	\hline
    $\theta=0$           & - & $-1/2$ & - & $3/8$ & - &$-5/16$  \\
	$\theta=1$           & $R$ & - & $-(9/8)R$ &- &   $({2375}/{2048})R$&-   \\
	$\theta=2$      &- & $R^2$ & - & -2$R^2$ &  - &$(351/128)R^2$  \\
	$\theta=3$           & - & - & $R^3$ & - &  $-(25/8)R^3$&-   \\
	$\theta=4$           & - & - & - & $R^4$ &  - &$-9/2 R^4$  \\
	$\theta=5$           & - & - & - & - &  $R^5$&-   \\
	$\theta=6$           & - & - & - & - &  -&$R^6$   \\
	\hline
\end{tabular}
	\caption{Coefficient of the universal term ($\log(R/\epsilon)$) which appear in reading the HEE for sphere in ${d+2}$ dimensional hyperscaling violating background. Note they appear up to the overall factor  of $\frac{\pi^{d/2}}{2G_N\Gamma(\frac{d}{2})}$. }
\end{table}
One observes that for odd $d$ and $\theta=0$, there is universal term which is given by
\be\label{universal} S^{\theta=0}_{uni}=(-1)^{\frac{d-1}{2}}\frac{\pi^{d/2}}{2G_N\Gamma(\frac{d}{2})}\frac{(d-2)!!}{(d-1)!!}\log(\frac{R}{\epsilon}), \ee
it is in agreement with \eqref{takaya}. For $\theta=d-1$ the log terms indicates that the corresponding background could provide a gravitational dual for a theory with a Fermi surface and the corresponding HEE is given by\footnote{Note that such theories have
	a dimensionful scale that does not decouple in the infrared and hence, dimensional analysis is restored identically  \cite{Dong:2012se}.}
\be S^{\theta=d-1}_{uni}=\frac{\pi^{d/2}}{2G_N\Gamma(\frac{d}{2})}R^{d-1}\log(\frac{R}{\epsilon}). \ee
The other universal $\log$ terms appear for $d\geq 3$ and $\theta=d-3$ moreover for $d\geq 5$ and  $\theta=d-5$ which we have computed them in table 1. It is worth to mention that in the absence of logarithms, finite contributions might be universal as well\footnote{We thank the referee for his/her comment on this point.}. For even $d$, e.g., for a 6-dimensional background we find the finite part of the HEE as follows
\be
S_{finite}=\frac{\pi^2}{4G_N}\frac{R^\theta(2\theta^2+\theta-12)}{(\theta-3)^2(\theta-1)},
\ee
where for $\theta=0$ one recovers the finite part of the HEE given by equation \eqref{takaya} for $d=4$.
\subsection{Cylindrical entangling region  }

For a cylinder, we parameterize the metric \eqref{lif2} as
\be\label{lif2cyl}
ds^2 =r^{\frac{2\theta}{d}-2} \Big(-r^{2(z-1)}dt^2+dr^2+d\rho^2+\rho^2d\Omega_{d-2}^2
\Big),\ee
where the entangling surface is a cylinder with $\rho = \ell$ on the $t = 0$ surface in this boundary geometry. We also introduce a regulator $H$ along the cylinder length. By taking the profile as $\rho=\rho(r)$, the entropy functional becomes
\be\label{cylinderee}
S=\frac{H}{2G_N} \frac{\pi^{\frac{d-1}{2}}}{\Gamma(\frac{d-1}{2})}\int dr  r^{ \theta -d}\,\rho(r)^{d-2} \sqrt{ \rho'(r)^2+1}.
\ee
One should minimize the above integrand to find the profile in the bulk. However, similar to the previous case we are interested in the universal terms, so what we need is actually to find the near boundary  behavior of the profile. By imposing the condition in which at the boundary $r=0$ one has $\rho(0)=\ell$ and after minimizing the entropy functional, near the boundary the profile takes the following form
\be
\rho(r)=\ell+\frac{(d-2)r^2}{2 (\theta-d +1)\ell}+\cdots.
\ee
Correspondingly the holographic entanglement entropy near the boundary becomes 
\be\label{cylinderee1}
S=\frac{H}{2G_N} \frac{\pi^{\frac{d-1}{2}}}{\Gamma(\frac{d-1}{2})}\int dr \, r^{\theta-d}\Big(\ell^{d-2}+\frac{(d-2)^2r^2\ell^{d-4}(\theta-d+2)}{2(\theta-d+2)^2}+{\cal O}(r^3)\Big).
\ee
We have computed the logarithmic  terms of the HEE in table 2. 
As one observes, similar to the previous case, for the special values of the hyperscaling violating exponent, the holographic entanglement entropy shows a logarithmic violation of the area law. The finite part of the HEE for a 6-dimensional background is given by 
\be
S_{finite}=\frac{\pi H R^{\theta
		-1}}{G_N}\frac{  \Big[  \Big(  ((\theta -11) \theta +42)-86\Big)\theta+181\Big]\theta-319 }{(\theta -5)^4 (\theta -3) (\theta -1)}.
\ee

\begin{table}[h!]
	
	\vspace{10pt}
	\centering
	\begin{tabular}{c|ccccc}
		\hline
		   & $d=3$ & $d=4$ & $d=5$ & $d=6$&$d=7$  \\
		\hline
		\hline
		$\theta=0$            &$ -(1/8\ell)$ & - & $(135/2048\ell)$ & - &$\sim0.025\frac{1}{\ell}$  \\
		$\theta=1$           & - & $-1/2$ & - &  $3/8$&-   \\
		$\theta=2$      & $\ell$ & - & $-(9/8)\ell$ &-&$\sim1.15\ell$  \\
		$\theta=3$            & - & $\ell^2$ & - & $-2\ell^2$ &-   \\
		$\theta=4$           & - & - & $\ell^3$ &  - &$-(25/8)\ell^3$  \\
		$\theta=5$            & - & - & - &  $\ell^4$&-   \\
		$\theta=6$            & - & - & - &  -&$\ell^5$   \\
		\hline
	\end{tabular}
\caption{Coefficient of the universal term ($\log(\ell/\epsilon)$) up to overall factor $\frac{H}{2G_N} \frac{\pi^{\frac{d-1}{2}}}{\Gamma(\frac{d-1}{2})} $ which appear in reading the HEE for cylinder in ${d+2}$ dimensional hyperscaling violating background.  }
\end{table}

In the following section we will add the higher order (Gauss-Bonnet) terms  into the Lagrangian and investigate  the logarithmic terms in a sphere and a cylinder as the entangling regions on the boundary in hyperscaling violating backgrounds.
\section{Holographic entanglement entropy in Hyperscaling violating backgrounds: Curvature corrections}
In this section we consider the HEE for systems with hyperscaling violation when the gravity action contains higher curvature terms. We  add the higher order (Gauss-Bonnet) terms  into the Lagrangian and investigate  the logarithmic terms of the HEE in a sphere and a cylinder as the entangling regions on the boundary in hyperscaling violating backgrounds. \\ 
As already mentioned in the previous section, one can use the RT-proposal to compute the HEE, however, this elegant picture does not work for actions with higher-derivative terms. There are some proposals to compute the HEE in higher curvature theories \cite{Buchel:2009sk, Hung:2011xb, deBoer:2011wk}. 
 We use the proposal of Ref. \cite{Fursaev:2013fta} where for the following action in the bulk with curvature squared terms \cite{Dong:2013qoa} 
\begin{equation}\label{action}
{\cal I} = \frac{1}{{16\pi {G_N}}}\int\limits_{\cal M} {{d^{d + 2}}x} \sqrt { - g} \left[ {R +\frac{{d(d+1)}}{{{L^2}}}  + {L}^2\Big(a_1{R^2} + a_2{R_{\mu \nu }}{R^{\mu \nu }} + a_3{R_{\mu \nu \rho \sigma }}{R^{\mu \nu \rho \sigma }} \Big) } \right],
\end{equation}
the HEE is given by 
\begin{multline}\label{fur1}
S = \frac{1}{{4{G_N}}} \int\limits_\Sigma {d^{d}x}\sqrt \sigma \bigg\{1+ {L}^2 \left[ {2a_1 R +a_2 \left( {{R_{\mu \nu }}n_i^\mu n_i^\nu  - \frac{1}{2}\sum\limits_i {{{\left( {Tr{\mathcal{K}^{\left( i \right)}}} \right)}^2}} } \right)} \right. \\
\left. { + 2a_3 \left( {{R_{\mu \nu \alpha \beta }}n_i^\mu n_i^\alpha n_j^\nu n_j^\beta  - \sum\limits_i {\mathcal{K}_{\mu \nu }^{\left( i \right)}\mathcal{K}_{\left( i \right)}^{\mu \nu }} } \right)} \right]\bigg\}.
\end{multline}
In the above relations ${L}$ is a length scale which for Einstein gravity reduces to AdS radius and $\Sigma$ is a co-dimension two hypersurface in the bulk spacetime ${\cal M}$ which extremizes the entropy functional. Note that introducing the curvature scale  ${L}$  makes the coupling constants to be dimensionless. 
The orthogonal normal vectors of $\Sigma $ are denoted by ${n_i}\,\left( {i = 1,2} \right)$ whereas the extrinsic curvature tensors are defined by 
\begin{equation}\label{extrinsic-curvature}
\mathcal{K}_{\mu \nu }^{\left( i \right)} = h_\mu ^\lambda h_\nu ^\rho {\left( {{n_i}} \right)_{\lambda ;\rho }},\,\,\,\,\,\,\,\,\,\,\,\,\,\,\,h_\mu ^\lambda  = \delta _\mu ^\lambda  + \xi \sum\limits_i {{{\left( {{n_i}} \right)}_\mu }{{\left( {{n_i}} \right)}^\lambda }} ,
\end{equation}
where $\xi$ is $+1$ for time-like and $-1$ for space-like vectors. It is noted that by setting the higher curvature coupling to zero, \eqref{fur1} reduces to the RT-proposal.

We will consider the Gauss-Bonnet (GB) gravity which can be achieved by setting
$a_1=a_3=-\frac{1}{4} a_2 = \frac{1}{(d-1)(d-2)}\lambda $ in \eqref{action} where  $\lambda$ is a dimensionless coupling constant that controls the strength of the GB term. The HEE for GB gravity has been studied in details in \cite{Hung:2011xb}. Also in hyperscaling violating background, Ref. \cite{Bueno:2014oua}  studied HEE for the strip.\footnote{The effects of Gauss-Bonnet corrections on some nonlocal probes in the holographic model with momentum relaxation has been studied in Ref. \cite{Tanhayi:2016uui}.} \\In the following of this section, we will first review the effect of curvature correction (GB term) on the HEE of a strip and then investigate the  logarithmically divergent terms of the HEE for a spherical and a cylindrical entangling surfaces.   

\subsection{Strip }

Similar to the Hilbert-Einstein gravity, in the case of the squared curvature gravity, by adding dilatonic scalar field with a nontrivial potential and a gauge field, one can extract a hyperscaling violating solution such as \eqref{lif2} \cite{Knodel:2013fua- Ghodrati:2014spa}. In this metric the following are the Ricci scalar and nonzero component of the Ricci tensor
 
 \begin{eqnarray}
 R&=&-\frac{2r^{-2\theta/d}}{\tilde{L}^2d}\bigg( z^2d+z\Big(d^2-(d+1)\theta\Big)+\frac{d+1}{2}(\theta-d)^2\bigg),\nonumber
 \\
 R_{tt}&=&\frac{r^{-2z}}{d}\bigg(z^2d+z\Big(d^2-(d+1)\theta\Big)+(\theta-d)\theta\bigg),\nonumber\\
 R_{rr}&=&\frac{-z^2+(\theta-d)+\frac{z\theta}{d}}{r^2},\nonumber\\
 R_{ii}&=&\frac{(d+z-\theta)(\theta-d)}{r^2d},\,\,\,\,\,i=1,\cdots,d.
 \end{eqnarray}
 In order to compute the HEE for strip, similar to the Einstein gravity, the profile can be parametrized as $x_1=x_1(r)$, so that the induced metric reads as
\begin{equation}
{ds}_{\mbox{ind}}^{2}=\tilde{L}^2 r^{\frac{2(\theta-d)}{d}}\Big[(1+{x'_1}^{2})dr^2+\sum_{i=2}^{d} dx_{i}^{2}\Big].
\end{equation}
After doing some calculation and making use of the relation \eqref{fur1}, the expression of HEE in the hyperscaling violating background when the gravity action contains higher order curvature terms is given by
\begin{multline}\label{fursaev1}
S = \frac{1}{{4{G_N}}} \int\limits_\Sigma r^{(\theta-d)}{\sqrt {1+{x'_1}^2}}\, {d^{d}x} \bigg\{1- \frac{{L}^2}{\tilde{L}^2} \left[ \frac{{a_1}r^{-2\theta/d}}{d(1+{x'_1}^2)^3}\Big(4z^2d+z(4d^2-(d+1)^2\theta)+2(d+1)(\theta-d)\Big) \right. \\
\left.{ +a_2(1+{x'_1}^2)^2\bigg(  (d+zd-2\theta)(d+z-\theta)+\Big(2z^2d+z\big(d^2+z(d^2-(d+2)\theta)\big)+(\theta-d)^2\Big){x'_1}^2}\right.\\
\left.{-\frac{d}{2}\Big((\theta-d)({x'_1}+{x'_1}^3)+rx_1''\Big)^2\bigg) + a_3\bigg( \frac{4}{d}(\theta-zd)(1+{x'_1}^2)^2\Big(\theta-d-zd{x'_1}^2\Big) }\right. \\
\left.{-r^2{x_1''}^2d+2(\theta-d)^2({x_1'}+{x_1'}^3)^2+4r(\theta-d)(x_1'+{x_1'}^3)x_1''}\bigg) \right]\bigg\}.
\end{multline}
Now one should find the profile of the hypersurface $x_1(r)$ which is supposed to be a smooth differentiable function with the condition $x_1(0) = \ell /2$. This is a complicated expression, nevertheless, this can be done by minimizing the entropy functional \eqref{fursaev1}. One may consider the entropy functional as a one-dimensional action by considering it as $S=\int dr {\cal L}$ in which the Lagrangian is independent of $x_1(r)$.  Therefore, the corresponding equation of motion becomes
\begin{equation}
\frac{\partial}{\partial r}(\frac{\partial {\cal L}}{\partial x_1''})-(\frac{\partial {\cal L}}{\partial x_1'})=C, \,\,\,\,\mbox{with}\,\,\,\,\frac{\partial {\cal L}}{\partial x_1}=0,
\end{equation}
 $C$ is a constant fixed by imposing the turning point of the hypersurface in the bulk where one has $x_1'(r_t)\rightarrow \infty$. Therefore, for the GB gravity one gets
\be
\frac{\tilde{L}^{-2}r^{\frac{(d-2)\theta }{d}-d} x_1' \Big(2 (\theta -d)^2L^2 \lambda -\tilde{L}^2{d^2} r^{2 \theta /d}
	\left(\left(x_1'\right){}^2+1\right)\Big)}{{d^2}
	\Big(\left(x_1'\right){}^2+1\Big){}^{3/2}}=-\frac{1}{r_t^{d-\theta}}. 
\ee
One can solve the equations of motion for small $\lambda$ and find the following expression for the profile 
\be\label{prof}
x'_1(r)=r^d r_t^{\theta }
\sqrt{\frac{1}{r_t^{2d} r^{2 \theta }-r^{2d} r_t^{2 \theta }}}+\frac{2}{d^2} (\theta -d)^2   r^{d-\frac{2 \theta }{d}} r_t^{\theta }
\sqrt{\frac{1}{r_t^{2d} r^{2 \theta }-r^{2d} r_t^{2 \theta }}}\,\,\lambda+{\cal O}(\lambda^2).
\ee
After making use of the relation $\ell=2\int_{0}^{r_t}x'_1(r)\,dr$, the following expression between the turning point $r_t$ and the width of the entangling region can be obtained  
\be
\ell=\frac{2 \sqrt{\pi } \Gamma \left(\frac{-d+\theta -1}{2 (\theta -d)}\right) r_t}{\Gamma
	\left(\frac{1}{2 d-2 \theta }\right)}-\frac{2\sqrt{\pi}(\theta-d)}{d^2}\frac{ \Gamma \left(\frac{d (d+1)-(d+2) \theta }{2 d (d-\theta )}\right)
	r_t^{1-\frac{2 \theta }{d}}}{ \Gamma \left(\frac{d (2 d+1)-2 (d+1) \theta }{2 d
		(d-\theta )}\right)} \lambda  +{\cal O}(\lambda^2).\label{ell}
\ee
 The equation \eqref{ell} can be inverted to find the turning point and by plugging the profile \eqref{prof} into the entropy function and after doing some calculation, one obtains the expression of the HEE in terms of the turning point. And finally this leads to HEE which is as follows
\be\label{EEGB}
S_{GB}=S_{EG}+\frac{\tilde{L}^d H^{d-1}}{2G_N}\bigg(\frac{\mathfrak{A}}{\epsilon^{-\frac{(d-2) \theta }{d}+d-1}}+\frac{\mathfrak{C}}{\ell^{-\frac{(d-2) \theta }{d}+d-1}}\bigg)\lambda+{\cal O}(\lambda^2),
\ee
where $S_{EG}$ stands for the HEE in Einstein gravity which is given by \eqref{EE}. $\mathfrak{A}$ , $\mathfrak{C} $ are two numerical constant and for $\theta=0$ and $\theta=1$ we have listed them in table 3. It is worth to mention that similar statement has been obtained in Ref. \cite{Bueno:2014oua} where the authors used proposal of Ref. \cite{Hung:2011xb} to compute the HEE of a strip in GB gravity. Noting that the GB gravity is a special form of curvature squared action, the final result of HEE deduced from the proposal of \cite{Hung:2011xb} (see also \cite{deBoer:2011wk}) would be the same with \cite{Fursaev:2013fta}, namely the universal parts of two computations coincide; However, taking into account the boundary terms would modify the coefficient of leading UV-divergent term. Regarding this fact, our result of HEE for a strip coincides with \cite{Bueno:2014oua}. 

	\begin{table}
		\centering
		\begin{tabular}{@{}l*2{>{\text\ttfamily}l}%
				l<{}l@{}l*2{>{\text\ttfamily}l}}
			\toprule[1.5pt]
			& \multicolumn{2}{c}{\head{$d+2=5$}}
			& \multicolumn{2}{c}{\head{$d+2=6$}}\\
				\cmidrule(lr){2-3}\cmidrule(l){3-5}
			& \normal{\head{$\theta=0$}} & \normal{\head{$\theta=1$}}
			& \normal{\head{$\theta=0$}} & \head{$\theta=1$}\\
			\cmidrule(lr){2-3}\cmidrule(l){4-5}
			\multirow{1}{*}{$ \mathfrak{A}$} & $-3$ & $-\frac{32}{15}$  & $-4$ & $-\frac{63}{20}$  \\\\
			\cmidrule(lr){2-3}\cmidrule(lr){4-5}
			\multirow{1}{1.1cm}{$ \mathfrak{C}$} & $-\frac{12 \pi ^{3/2} \Gamma \left(\frac{2}{3}\right)^3}{\Gamma
				\left(\frac{1}{6}\right)^3}$ & $-
			\frac{ 2^{8/3} \pi ^{4/3} \Gamma \left(-\frac{5}{12}\right) \Gamma
			\left(\frac{3}{4}\right)^{5/3}}{9 \Gamma
			\left(\frac{1}{12}\right)\Gamma \left(\frac{1}{4}\right)^{5/3}}$  & $-\frac{64 \pi ^2 \Gamma \left(\frac{5}{8}\right)^4}{\Gamma \left(\frac{1}{8}\right)^4}$ & $-\frac{9 \pi ^{11/4} \Gamma \left(\frac{1}{3}\right) \Gamma
			\left(\frac{2}{3}\right)^{3/2}}{10 \Gamma \left(\frac{1}{6}\right)^{5/2} \Gamma
			\left(\frac{13}{12}\right)^2}$ 
			& \multirow{2}{1.8cm}{\sffamily\bfseries  } \\\\
					\bottomrule[1.5pt]
		\end{tabular}
		\caption{The numerical constant terms appear in the computation HEE of a strip in five and six dimensional hyperscaling violating backgrounds. }
	\end{table}

It is clear that in computing the HEE in the static background of GB gravity, dynamical exponent $z$ does not play any role. Moreover, it can be seen that in any case at the zero order of the GB parameter, $\epsilon$ and $\ell$ scale as the power of $d-\theta-1$ whereas at the first order of $\lambda$ they scale as $d-\theta+\frac{2\theta}{d}-1$.


\subsection{spherical and cylindrical entangling regions}
In the case of spherical entangling region, similar to the Einstein gravity, the entangling region on the boundary is given by $r<R$ where at  $t=0$. Therefore,  the co-dimension two surface in the bulk is parametrized by $\rho=\rho(r)$ and in this case, the functional of the holographic entanglement entropy becomes 

$$
S=\frac{\tilde{L}^d}{2G_N}\frac{\pi^{d/2}}{\Gamma(\frac{d}{2})}\int   r^{ \theta -d}\rho^{d-1} \sqrt{ \rho'^2+1}\,dr \bigg[1+\frac{2L^2\lambda}{\tilde{L}^2d^2(d-1)(d-2)\rho^2(1+\rho'^2)^2}r^{-2\theta/d}\bigg((1-d)(1+\rho'^2)
$$
\be\label{shere1} \Big((\theta-d)((d-2)\theta-d^2)\rho^2+d(d+4)r(\theta-d)\rho\rho'+d^2r^2\rho'^2\Big)+d(d+4)r\rho\big((\theta-d)\rho\rho'-rd\big)\rho''\bigg)\bigg].\ee
Similar to the previous sections, one can consider the above entropy functional as an action, thus one just need to solve the equations of motion coming from the variation of the action to find the profile. However, as long as, we want to address the universal part of the HEE, it is only needed to write the near boundary behavior of the hypersurface in the bulk.  We found the expansion of the profile near the boundary $(r\approx 0)$ as follows 
\be
\rho(r)=R+\frac{d-1}{2R(\theta-d+1)}r^2-\frac{2\theta^2}{Rd(\theta-d+1)}\frac{ r^{2-2\theta/d}}{(d-2)\theta-d(d-1)}\lambda  +{\cal O}(\lambda^2).
\ee  
After doing some calculations, we have summarized the logarithmic terms of HEE up to first order of the coupling constant for the sphere entangling region in table 4. 
\begin{table}[h!]
	
	\vspace{20pt}
	\centering
	\begin{tabular}{|c|c|c|c|c|}
		\hline
		   & $d=3$ & $d=4$ & $d=5$ & $d=6$ \\
		\hline
		\hline
		$\theta=0$            & $-\frac{1}{2}+\frac{15}{4}\lambda$ & - & $\frac{1}{16}(6-35\lambda)$ & -  \\\hline
			$\theta=1$            & - & $\frac{9}{8}R(-1+2\lambda)$ & - &  $\sim1.15R(1-3\lambda)$   \\\hline
		$\theta=\frac{3}{2}$     & - & - & - &  $\sim-0.94\,\,R\lambda$  \\\hline
		$\theta=2$       & $R^2(1-\frac{3}{2}\lambda)$ & $\frac{3}{2}R\lambda$ & $R^2(-2+5\lambda)$ &  -  \\\hline
		$\theta=3$            & - & $R^3(1-2\lambda)$ & - &  $\frac{25}{8}R^3(-1+3\lambda)$  \\\hline	$\theta=\frac{10}{3}$            & - & - & $\frac{4}{3}R^2\lambda$ &  -  \\\hline
		$\theta=4$            & - & - & $R^4(1-\frac{5}{2}\lambda)$ &  -  \\\hline
		$\theta=\frac{9}{2}$    & - & - & - &  $\frac{5}{8}R^3\lambda$   \\\hline
		$\theta=5$            & - & - & - &  $R^5(1-3\lambda)$   \\\hline
		$\theta=6$            & - & -  &  - &- \\\hline
		
	\end{tabular}
\caption{Coefficient of the universal term ($\log(R/\epsilon)$) which appear in reading the HEE for sphere in ${d+2}$ dimensional hyperscaling violating background. The coefficients appear up to the overall factor  of $\frac{\pi^{d/2}\tilde{L}^d}{2G_N\Gamma(\frac{d}{2})}$.  }
\end{table}

We would like to mention that though we mainly consider a theory with hyperscaling violation, however, when it comes to the
comparison, we restrict ourselves to $\theta = 0$. 
In this way in Ref. \cite{Hung:2011xb-Myers:2012ed-Safdi:2012sn},  the universal term of the entanglement entropy in even-dimensional CFTs was obtained as follows 
\begin{equation}\label{univ}
S_{uni}=(-1)^{\frac{d-1}{2}}\frac{\pi^{\frac{d-1}{2}}\tilde{L}^d}{2G_N\Gamma(\frac{d+1}{2})}\Big(1-\frac{2d}{d-2}f_{\infty}\lambda\Big)\log(\frac{R}{\epsilon}),
\end{equation} 
where $f_\infty$ is given by 
\begin{equation}\label{finf}
\tilde{L}^2=\frac{L^2}{f_{\infty}},\,\,\,\,\,\,\,\,\,f_\infty=\frac{1-\sqrt{1-4\lambda}}{2\lambda}.
\end{equation} 
We note that at the linear level of the coupling constant $\lambda$,  our results match the \eqref{univ} for $\theta=0$. 




For a cylinder with $\rho = \ell$ at $t = 0$ surface which is defined on the geometry \eqref{lif2cyl}, we found the following expression for the holographic entropy functional  

\begin{multline}\label{cylinder1}
S=\frac{H\tilde{L}^{d-1}}{2G_N} \frac{\pi^{(d-1)/2}}{\Gamma(\frac{d-1}{2})}\int   r^{ \theta -d}\,\rho^{d-2} \sqrt{ \rho'^2+1}\,\,dr\bigg\{1-\frac{2L^2\lambda}{\tilde{L}^2d^2(d-1)(d-2)\rho^2(1+\rho'^2)^2}r^{-2\theta/d}\\
\Big[(1+\rho'^2)\Big((d-1)(\theta-d)((d-2)\theta-d^2)\rho^2+2d(d-1)(d-2)(\theta-d)r\rho\rho'+d^2(d-2)(d-3)r^2\rho'^2\Big)\\+2r d\Big(d(d-2)r-(d-1)(\theta-d)\rho\rho'\Big)\rho\rho'' \Big] \bigg\}.
\end{multline}
Note that in the bulk the profile is defined by $\rho=\rho(r)$ and the prime stands for the derivative with respect to $r$. The entropy functional \eqref{cylinder1} should be minimized and after making use of the similar arguments of the sphere, it is only needed to write the near boundary behavior of the hypersurface in the bulk.  We have summarized the coefficients of the logarithmic terms in table 5.

\begin{table}[h!]\label{cylinderlam}

	\vspace{20pt}
	\centering
	\begin{tabular}{|c|c|c|c|c|}
		\hline
		   & $d=3$ & $d=4$ & $d=5$ & $d=6$ \\
		\hline
		\hline
		$\theta=0$    & $\frac{1}{8\ell}(-1+2\lambda)$ & - & $\frac{45}{2048\ell}(3-2\lambda)$ & -  \\\hline
		$\theta=1$           & - &$-1/2$ & - &  $3/8$   \\\hline
		$\theta=\frac{3}{2}$    & - &- & - &  $\sim-0.147\,\lambda$   \\\hline 
		$\theta=2$       & $\ell$ & $\frac{-2}{3}\lambda$ & $-(9/8)\ell$ &  -  \\\hline
		$\theta=3$            & - & $\ell^2$ & - &  $-2\ell^2$  \\\hline
		$\theta=\frac{10}{3}$       & - & - & $\frac{3}{4}\ell\lambda$ &  -  \\\hline
		$\theta=4$           & - & - & $\ell^3$ &  -  \\\hline
		$\theta=\frac{9}{2}$   & - & - & - &  $\frac{2}{5}\ell^2\lambda$   \\\hline
		$\theta=5$           & - & - & - &  $\ell^4$   \\\hline
		$\theta=6$            & - & - & - &  -  \\\hline
		
	\end{tabular}
	\caption{Coefficient of the universal term ($\log(\ell/\epsilon)$) up to overall factor  $\frac{\tilde{L}^{d-1}H}{2G_N} \frac{\pi^{\frac{d-1}{2}}}{\Gamma(\frac{d-1}{2})} $  which appear in reading the HEE for cylinder in ${d+2}$ dimensional hyperscaling violating background. }
\end{table}


\section{Mutual and Tripartite Informations in Hyperscaling Violating Backgrounds}

It is already mentioned that entanglement entropy could potentially address some information about the underlying conformal field theory. However, appearance of the UV cutoff in the expression of entanglement entropy makes it a scheme-dependent quantity, so that it is natural to explore other measurement of entanglement to probe the field content of theory. For example by introducing an appropriate
linear combination of entanglement entropies, one can define an important quantity named as mutual information where for two disjoint regions say as $A$ and $B$, it is defined by 
 \be\label{mutu}
I(A:B)=S_A+S_B-S_{ {A\cup B}}, \ee where $S_i$'s stand
for the entanglement entropies of the corresponding region $i$ with the rest of system. It was shown that mutual information is a finite and positive valued quantity which measures the amount of entanglement shared between two systems $A$ and $B$. In other words mutual information quantifies the amount of information that we can obtain from one of the subsystems by looking at the other one and for two uncorrelated systems the mutual information becomes identically zero.\footnote{It is worth to mention that in Ref. \cite{Mozaffar:2015xue} it was shown that the
holographic	mutual information is divergent for singular entangling regions. The time evolution of such measures can be found for example in  \cite{Tanhayi:2015cax, Erdmenger:2017gdk, Flory:2017xqr,  Kundu:2016cgh, Hajilou:2017sxf }.}  In a 2-dimensional conformal field theory,  mutual information depends on the full operator content of the theory which indicates that this entanglement measure could carry more information about the underlying conformal field theories \cite{Calabrese:2009ez}. 

In the context of gauge-gravity duality by making use of the RT-proposal for computing HEE, we investigate some particular properties of mutual information between two disjoint rectangular systems in hyperscaling violating backgrounds. In order to study the mutual information, we compute this quantity for two disjoint strips where their lengths and separation are given by $\ell_1$, $\ell_2$ and $h$, respectively. In this case, as mentioned the HEE is proportional to the minimal area in the bulk and in computation of the entanglement entropy for the union of two entangling  strips namely $S_{{A\cup B}}$,  there are always two candidate minimal area surfaces in the bulk and depending on the separation of $A$ and $B$ one of them is indeed the minimum one \cite{Fischler:2012ca, Fischler:2012uv}. Therefore one can write the entanglement entropy for the union of two entangling strips as follows 
\bea\label{SAUB}
S_{{A\cup B}}=\Bigg\{ \begin{array}{rcl}
	&S(\ell_1+\ell_2+h)+S(h)&\,\,\,h< \ell_1, \ell_2,\\
	&S(\ell_1)+S(\ell_2)&\,\,\,h> \ell_1, \ell_2,
\end{array}\,\,
\eea
where $S(l)$ stands for the entanglement entropy of a strip
with width $l$. Thus the mutual information becomes
\bea\label{HMI}
I(A: B)=\Bigg\{ \begin{array}{rcl}
	&S(\ell_1)+S(\ell_2)-S(\ell_1+\ell_2+h)-S(h)&\,\,\,h< \ell_1, \ell_2,\\
	&0&\,\,\,h> \ell_1, \ell_2.
\end{array}\,\,
\eea
For two strips with the same width and after making use the above relation and also \eqref{mutu}, it is obvious that the mutual information becomes zero in the
case of $h> \ell$, on the other hand for $h< \ell$, one finds
\be
\label{MI} I(A:B)=2S(\ell)-S(2\ell+h)-S(h)\equiv I. 
\ee
In what follows we will also consider $h< \ell$. This indicates
that finding the mutual information of two parallel
strips is identical to the entanglement entropy of three strips  with widths $h, \ell$ and $2\ell+h$. We consider the case in which the widths of two parallel strips are same, otherwise one has to compute four
entanglement entropies corresponding to $\ell_1, \ell_2, h $ and
$\ell_1+\ell_2+h$.\\
The entanglement
entropy for a strip in a $d+1$ dimensional CFT whose gravity dual
is provided by the hyperscaling violating geometry,  for Einstein and GB gravities are given by \eqref{EE} and \eqref{EEGB}, respectively. In the case of GB gravity in AdS$_5$ background, one obtains

\be
I_{GB}=\frac{H^2}{4G_N(2-\theta)}\bigg[\mathfrak{A}\bigg(\frac{-2}{\ell^{2-\theta}}+\frac{1}{(2\ell+h)^{2-\theta}}+\frac{1}{h^{2-\theta}}\bigg)+\mathfrak{C}\,\,\lambda\bigg(\frac{-2}{\ell^{2-\theta/3}}+\frac{1}{(2\ell+h)^{2-\theta/3}}+\frac{1}{h^{2-\theta/3}}\bigg)\bigg],
\ee
where $\mathfrak{A}$ and $\mathfrak{C}$ are two numerical constants given in table 3.

The above consideration can be easily generalized to
systems consisting of $n$ disjoint subsystems. If one interested in the entanglement entropy for a system consists of
$n$ disjoint regions
$A_i,\;i=1,\cdots, n$, then a new quantity should be defined which is known as the $n$-partite information. As a matter of fact this new quantity examines the role of measurement of the amount of information or correlations (both classical and quantum) between those subsystems. In terms of entanglement entropy one may define the $n$-partite
information as follows \cite{Hayden:2011ag}\footnote{It is worth to mention that this definition is not unique; For other generalization see \cite{Horodecki:2009zz}.
} \be\label{JAB}
I^{[n]}(A_{\{i\}})=\sum_{i=1}^nS(A_i)-\sum_{i<j}^n S(A_i\cup
A_j)+\sum_{i<j<k}^n S(A_i\cup A_j\cup A_k) -\cdots\cdots -(-1)^n
S(A_1\cup A_2\cup\cdots\cup A_n), \ee where similar to the mutual information, $S(A_i\cup A_j\cdots
)$ is the entanglement entropy of the region $A_i\cup A_j\dots$
with the rest of the system. Note that for $n=1$ and $n=2$ this formula gives us entanglement entropy and mutual information. The case of $n=3$ which is known as the three-partite (tripartite) information is given by
\begin{align}\label{3par}
{I^{[3]}}\left( {{A_1}:{A_2}:{A_3}} \right) &= S\left( {{A_1}} \right) + S\left( {{A_2}} \right) + S\left( {{A_3}} \right) - S\left( {{A_1} \cup {A_2}} \right) - S\left( {{A_1} \cup {A_3}} \right) \notag\\
&- S\left( {{A_2} \cup {A_3}} \right) + S\left( {{A_1} \cup {A_2} \cup {A_3}} \right).
\end{align}  
It was argued that holographic tripartite information has a fixed sign it is always negative \cite{Hayden:2011ag, Alishahiha:2014jxa}, this is indeed the negativity of tripartite information which results in an important inequality relation on the mutual information. To explicitly see this constraint let us write \eqref{3par} in terms of the mutual information as
\begin{align}\label{3par1}
{I^{[3]}}\left( {{A_1}:{A_2}:{A_3}} \right){\rm{ }} = I\left( {{A_1} : {A_2}} \right) + I\left( {{A_1} : {A_3}} \right) - I({A_1}:{A_2} \cup {A_3}),
\end{align} 
which the negativity of tripartite information leads one to write
\begin{align}\label{3pa}
I({A_1}:{A_2} \cup {A_3})\ge I\left( {{A_1} : {A_2}} \right) + I\left( {{A_1} : {A_3}} \right).
\end{align} 
In the context of quantum information theory this kind of inequality known as monogamy relation of mutual information. It is an important characteristic of measures of quantum entanglement which states that correlations in holographic theories arise primarily from entanglement rather
than classical correlations. The above constrain suggests that  if two systems say as $A_1$ and $A_2$ with the same finite Hilbert space dimension are maximally entangled, then neither of them can share any correlation with a third system $A_3$ \cite{Lancien:2016nvr}.\footnote{To investigate other inequalities on holographic entanglement measures see \cite{Erdmenger:2017gdk}. }

Now the main aim is to investigate the sign of these measures in hyperscaling violating theories. For two intervals of the same length $\ell$ separated by distance $h$, by semi-analytic computation we show that in all range of parameters that we have considered the mutual information and tripartite information are always positive and negative, respectively, which means the inequality \eqref{3pa} still holds. Therefore the mutual information is monogamous in hyperscaling violation theories. The following figures demonstrate this fact for hyperscaling violating background in five dimensional background.

\begin{figure}[h]
	\centering
	\includegraphics[width=7.3cm]{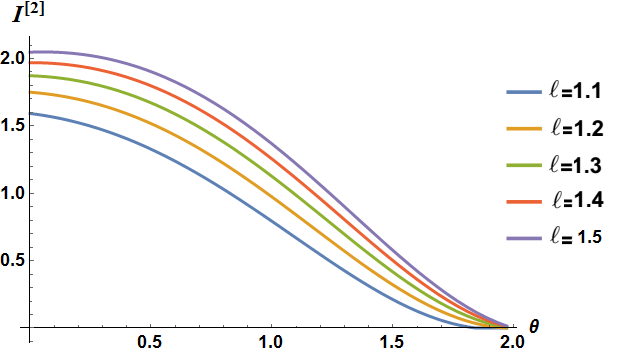}\,\,\,\,\,\,\,\includegraphics[width=7.3cm]{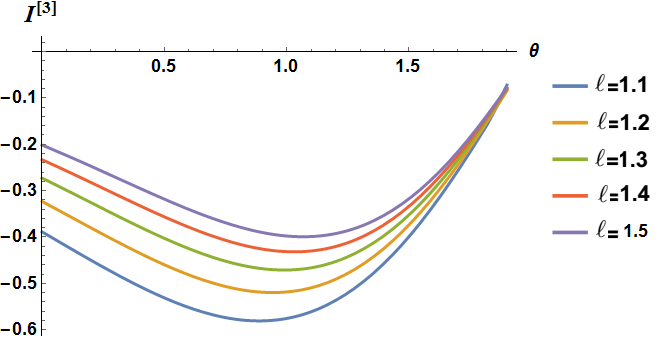}
	\caption{Holographic mutual information (\emph{left}) and tripartite information (\emph{right}) in $5-$dimensional hyperscaling violating background for $h =0.5,\,\,\, \lambda=0$ and different values of entangling width $\ell=1.1,1.2,1.3,1.4$ and $\ell=1.5$.}\label{n2ads}
\end{figure}
\begin{figure}[h]
	\centering
	\includegraphics[width=7.2cm]{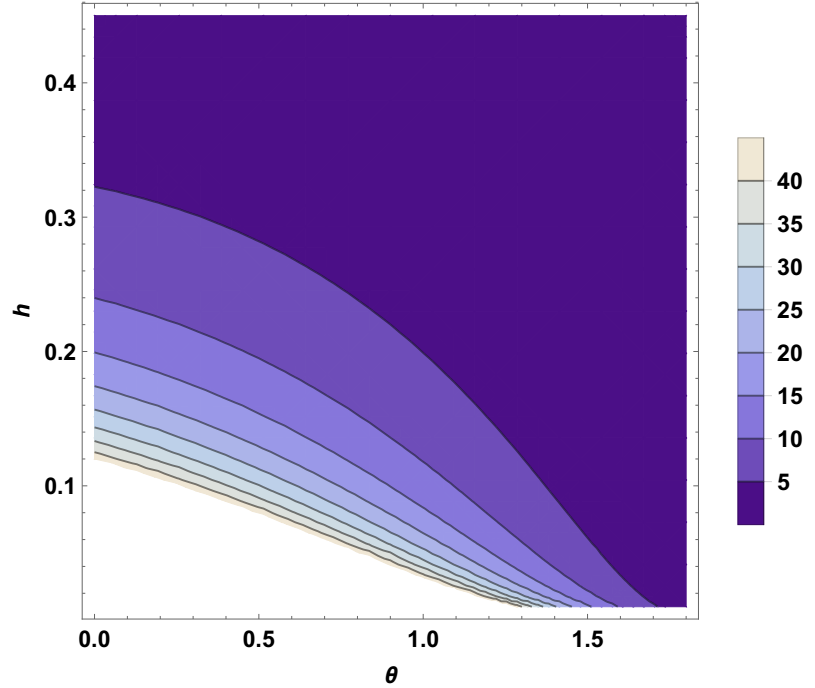}\,\,\,\,\,\,\,\includegraphics[width=7.2cm]{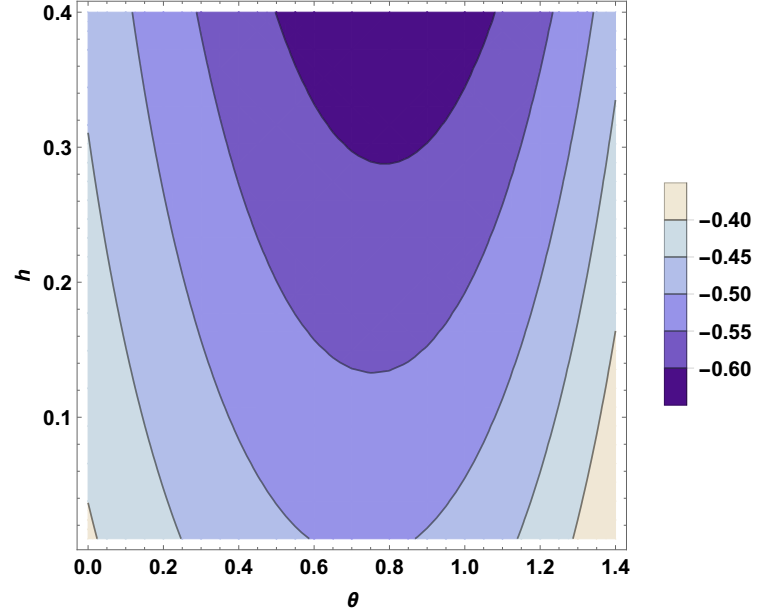}
	\caption{ Contour plot for mutual information $(left)$ and tripartite information $(right)$ in $5-$dimensional hyperscaling violating background  for $\ell=1,\,\,\,\,\lambda=0$ and for different values of $\theta$ and $h$.}
\end{figure}

\begin{figure}[h]
	\centering
	\includegraphics[width=7.3cm]{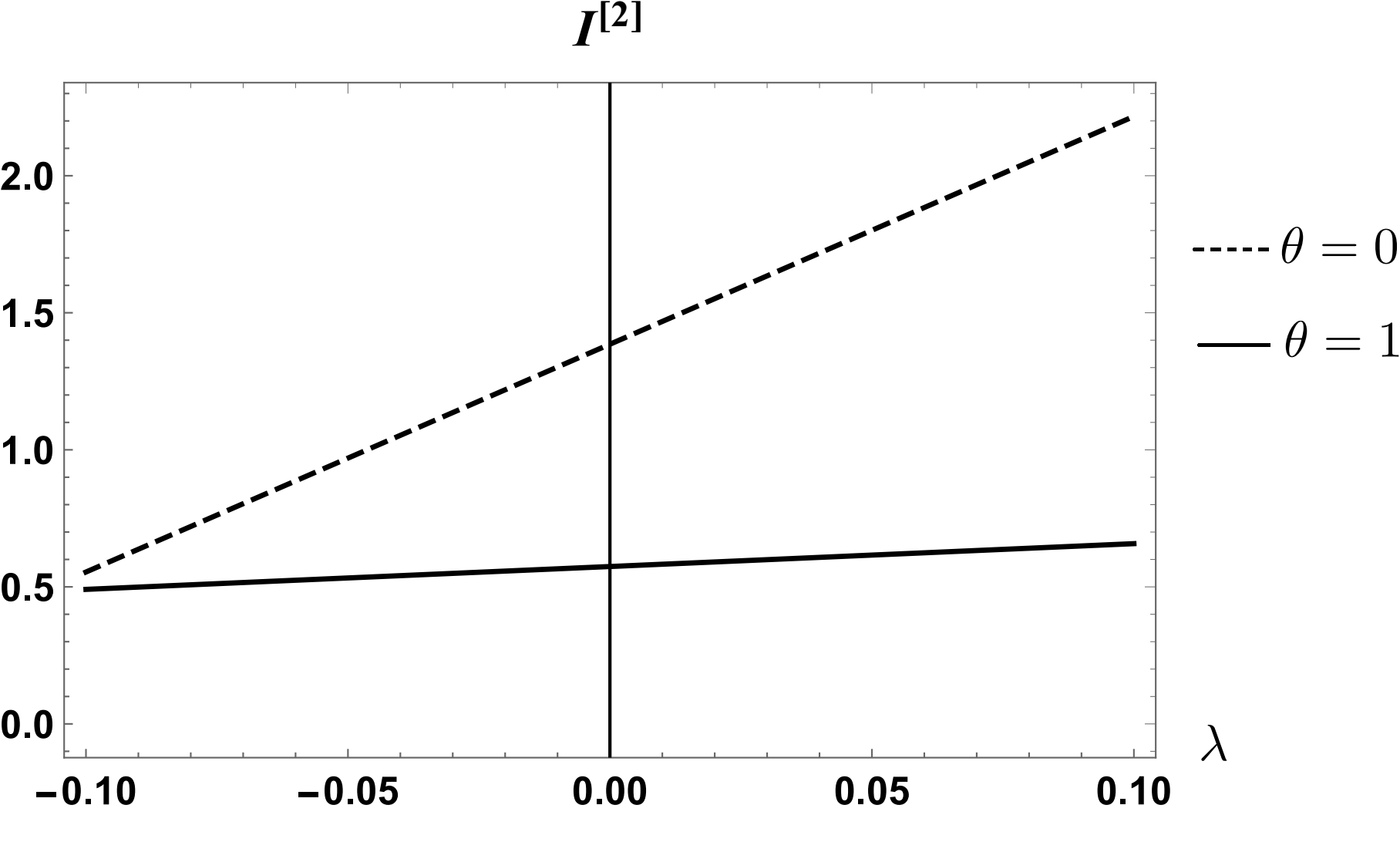}\,\,\,\,\,\,\,\includegraphics[width=7.3cm]{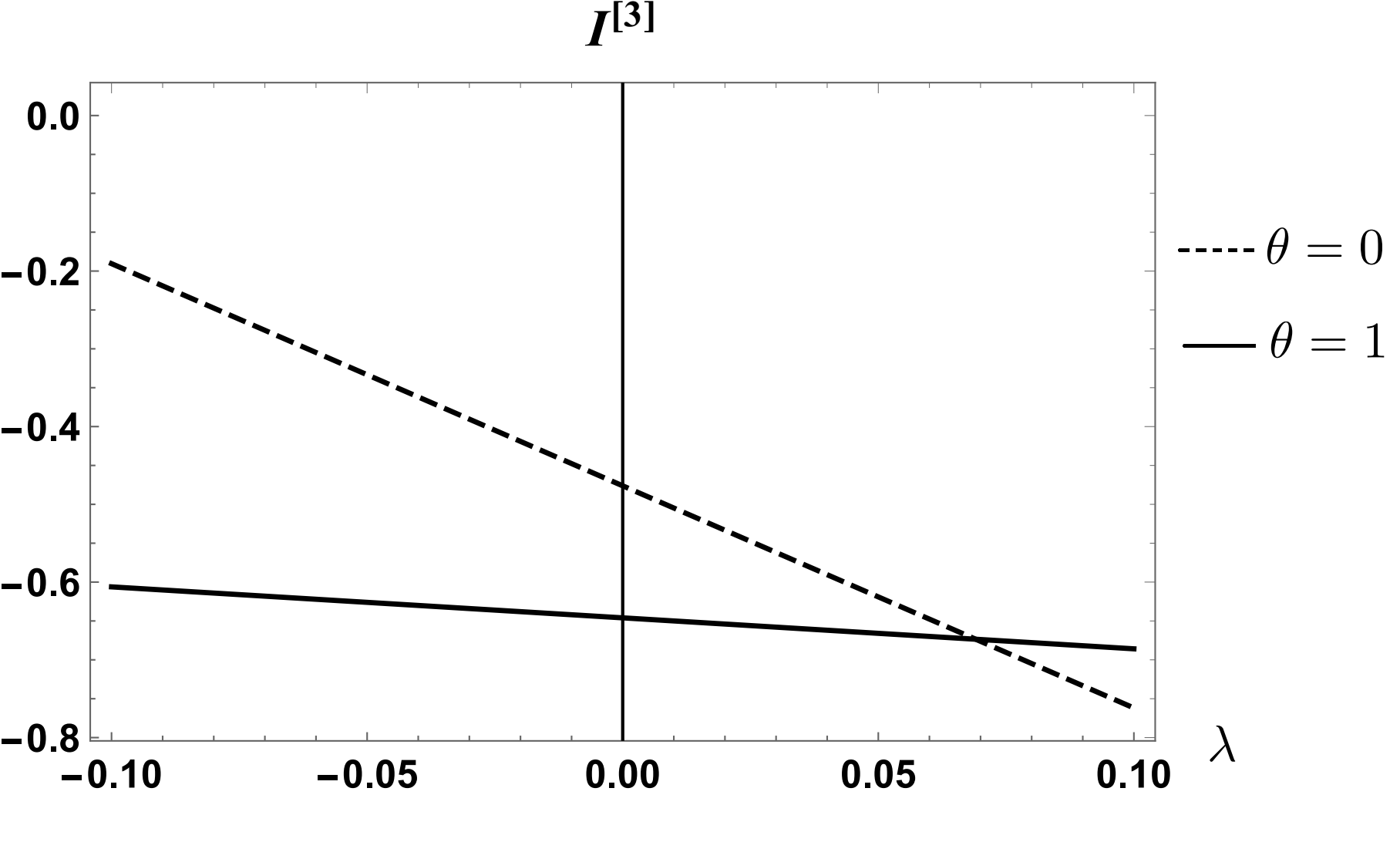}
	\caption{Holographic mutual information (\emph{left}) and tripartite information (\emph{right}) in $5-$dimensional  hyperscaling violating background with GB term. The graphs
		show different values of GB coupling for $\theta=0$ (dashed line) and $\theta=1$ (solid line) where in both cases we set $\ell=1$ and $h=0.5$.  }
\end{figure}



\section{Conclusion}
In this paper by making use of AdS/CFT correspondence, we obtained the universal terms of entanglement entropy for strongly coupled theories with hyperscaling violation. Previously it was shown that for $\theta=d-1$ and any dynamical exponent $z$,  the entanglement entropy receives a logarithmic term that indicates a violation of the area law \cite{{Alishahiha:2012cm}}.  In this work we observed that due to the hyperscaling parameter, the divergence structure of the entropy gets new terms. Explicitly in a $d+1$ dimensional field theory the logarithmically divergent terms appear for  $\theta=d-n$ in which $d\ge n,\,\,\,n=1,3,5,\cdots$. Keeping in mind the  dimensional analysis of such theories, the coefficient of the log term in the holographic entanglement entropy scales as $R^{d-n}$ for a sphere and $\ell^{d-n-1}$ for a cylinder. On the other hand, when the theory includes higher-curvature terms some other new logarithmic terms appear in HEE and also  the universal terms of the entanglement entropy get corrected non-trivially where in the case of GB terms, we obtained these corrections.  

We also examined the mutual  information in this setup for two parallel strips with width $\ell$ separated by distance $h$ assuming that $h<\ell$. Actually for two separated systems,  the mutual information is a measure of quantifying the amount of entanglement (or information) that these two systems can share. Mutual information is monogamous in a sense that it obeys the following inequality  $$I({A_1}:{A_2})+I({A_1}:{A_3})\leq I({A_1}:{A_2}\cup{A_3}).$$  
The above inequality states  that entangled correlations between $A_1$ and $A_2$ cannot be shared with a third system $A_3$ without spoiling the original entanglement \cite{Hayden:2011ag}. In general, it is a difficult task to identify the monogamy relation, however in principle, the holography methods provides a framework to study above inequality by looking at the tripartite information. The expression of tripartite information is given by \eqref{3par1} and its negativity leads to the above mentioned monogamy of mutual information. Using the holographic methods, in hyperscaling violating theories we showed that the tripartite information is always negative and hence, mutual information is monogamous.

It is worth mentioning that investigating the entanglement inequalities which have been imposed by holographic prescriptions may be important in a sense that this 
may help us to distinguish between two quantum theories: those can have a holographic dual from the others that cannot. This might thought of as an intrinsic property of a field theory with gravity dual. One of the most important inequalities is the strong subadditivity which states for any three subsystems $A_1$, $A_2$ and $A_3$ that do not intersect each other, one has \cite{Lieb} 
$$ S(A_1\cup A_2)+S(A_2\cup A_3)-S(A_1\cup A_2\cup A_3)-S(A_2)\ge0,$$
by setting $A_2$ empty one obtains the subadditivity inequality $ S(A_1)+S(A_2)-S(A_1\cup A_2)\ge0,$ which leads to the definition of mutual information.  It is known that this inequality is also satisfied by using the holographic RT-proposal of finding HEE. Intuitively it would be interesting to generalize the above inequality to $n$ disjoint subsystems. When we are interested in
measurement of the amount of information or correlations (both
classical and quantum) between $n$ disjoint regions
$A_i,\;i=1,\cdots, n$, the $n$-partite information may provide a
good measure for studying this process. Making use of the relation \eqref{JAB} for $n$ intervals, previously we showed that for the special case where all intervals have the same length and distance from their neighboring systems one has \cite{Alishahiha:2014jxa, Mirabi:2016elb} 
$$I^{[4]}(A_1:A_2:A_3:A_4)\ge0,\,\,\,\,\mbox{and}\,\,\,\,I^{[5]}(A_1:A_2:A_3:A_4:A_5)\le0.$$
Though these inequalities do not hold in general holographic setups \cite{Erdmenger:2017gdk, Flory:2017xqr}, we observed that for the above mentioned special case of the same width and separation in hyperscaling violating backgrounds, they are still satisfied. \\As a future work one can  consider the time-dependent hyperscaling violating backgrounds and check
how the sign of these quantities changes during the thermalization process.

Moreover, for an even-dimensional CFT, the coefficient of
the logarithmic term in the entanglement entropy gives us the central charges and they can be determined through the trace anomaly. This ‘geometric approach’ was established for two and higher (even-) dimensional CFT's (\cite{Hung:2011xb-Myers:2012ed-Safdi:2012sn} and references therein). Thus, it would be interesting to investigate  the geometric root of the new logarithmically divergent terms found in this paper.


\section{Acknowledgments}

 I would like to thank Mohsen Alishahiha for the related comments. Without discussions with M. Reza
 Mohammadi Mozaffar and Ali Mollabashi this paper would probably not exist, hereby I also thank them.  I would also like to acknowledge  A. Naseh, F. Omidi, R. Vazirian, A. Akhavan, M. Vahidinia, A. Shirzad and M. Hasanloo for some related discussions.  We thank a very conscientious referee for offering several useful remarks. This work has
been supported in parts by Islamic Azad University Central Tehran
Branch.


\end{document}